# Sensing solvents with ultra-sensitive porous poly(ionic liquid) actuators

*Qiang Zhao,[a] Jan Heyda,[b,e] Joachim Dzubiella,[b,c] Karoline Täuber,[a] John W.C. Dunlop[d] and Jiayin Yuan*[a]*

[a]   Dr. Q. Zhao, Miss K. Täuber, Dr. J. Yuan
      Department of Colloid Chemistry, Max Planck Institute of Colloids and Interfaces, D-14424 Potsdam, Germany
      E-mail: jiayin.yuan@mpikg.mpg.de

[b]   Dr. J. Heyda, Prof. J. Dzubiella
      Soft Matter and Functional Materials, Helmholtz-Zentrum Berlin, Hahn-Meitner Platz 1, 14109 Berlin, Germany

[c]   Prof. J. Dzubiella
      Institut für Physik, Humboldt-Universität zu Berlin, Newtonstr. 15, 12489 Berlin, Germany.

[d]   Dr. John W.C. Dunlop
      Department of Biomaterials, Department, Max Planck Institute of Colloids and Interfaces, D-14424 Potsdam, Germany

[e]   Dr. J. Heyda
      Department of Physical Chemistry, Institute of Chemical Technology, Institution, Prague, Technická 5, 166 28 Praha 6, Czech Republic

Currently, synthetic polymer actuators are being actively pursued owing to their importance in artificial muscles, molecular motors, soft robotics, programmable origami, and energy generators.[1] These smart materials are capable of adaptive motion, and/or reversible shape variation responding to external stimuli.[2] Among various aspects of actuators, it is necessary to develop higher sensitivity in actuating setups to realize signaling output at a rather early stage of external trigger, which is important for both fundamental research and practical applications. In this regard biological actuators are compelling models possessing the defining ability to sense and respond to subtle alterations in environmental conditions such as humidity and forces.[3] For example, wheat awns can propel their seeds on and into the ground in response to humidity changes.[3d] The study on such biological actuators demonstrates the important role of material architecture (pore-size, fiber orientation, etc.) on actuation and highlights its potential in the design of artificial actuators. Recently, enormous efforts are being paid to make synthetic polymer actuators more sensitive, such as reducing the electric voltage for driving polymer electrolyte actuators,[4] decreasing the energy consumption for light-responsive actuators,[5] and improving the humidity sensitivity of hygroscopic actuators.[6] Despite much success achieved so far, there is still plenty of potential to improve the sensitivity of synthetic polymer actuators.

Solvent stimulus polymer actuators (SSPAs) represent an important mechanism encompassing a significant breadth of utility including responsive gels,[7] grippers,[8] nano-robotics,[9] and solvent stimulus shape memory polymers.[10] For SSPAs usually the solvent diffusion into actuators results in heterogeneous volume changes giving rise to macroscopic shape changes and/or adaptive movements. In some cases, solvents may not trigger the actuation directly; instead, the real stimuli (e.g., pH, electrolytes, etc.) are coupled to solvent systems and reach the site of action through the diffusion of solvents and swelling in the polymer matrix.[11] Yet like most polymer actuators, SSPAs suffer from a relatively low sensitivity—usually a substantial amount of secondary solvents is required to mix with the primary solvent in order to produce noticeable shape deformation or displacement.[12] In some cases SSPAs were even shuttled between two different solvents to acquire the response, because a large gradient in solvent concentration is a necessity for promoting the solvent diffusion into bulk polymers.[12a] Thus large-scale actuation triggered by a low portion of solvent stimulus (e.g., < 0.5 mol%) remains an elusive challenge.

Here we report a porous poly(ionic liquid) (PIL) membrane actuator exhibiting exceptional sensitivity to low organic solvent concentrations. The membrane preparation follows a similar method we recently invented for porous polymer actuators in gas phase.[13]

Since solution phase actuation is more general for polymer actuators, our interest was directed to construct new SSPAs with high sensitivity. The actuator readily bends to an arc (curvature 0.076 mm$^{-1}$) upon adding as low as 0.25 mol% acetone molecules (1 acetone per 400 water molecules). To make a quantitative comparison, we define the actuator's sensitivity to organic solvents concentration as the amount of curvature change triggered by adding 1 mol% of the solvents. Thus our membrane is found to be at least one order of magnitude more sensitive than other state-of-the-art SSPAs.

The membrane chemically consists of two polyelectrolytes, a cationic PIL, poly[3-cyanomethyl-1-vinylimidazolium bis(trifuoromethanesulfonyl)imide] (denoted as PCMVImTf$_2$N), and poly(acrylic acid) (denoted as PAA, molecular weight 2 000 Da, a commercial product from Sigma Aldrich). Specifically speaking, PILs are polymerization products of ionic liquid monomers. The high density packing of ionic liquid species in PILs gives rise to distinctive properties, e.g., tunable solubility in organic media, surface activities, broad glass transition temperature, etc.[14] Recently there is huge attention on applying PILs as innovative polyelectrolytes to build up advanced materials and (multi)functional devices.[15] The PCMVImTf$_2$N used in this research was synthesized according to our previous report (Figure S1, supporting information).[16]

To prepare the porous membrane, PCMVImTf$_2$N and PAA were dissolved in dimethylformamide, solution-cast on a glass plate, dried (80 °C, 2 h), and subsequently soaked in aqueous ammonia (0.2 wt%, 20 °C, 2 h). Afterwards a free-standing membrane (denoted as PCMVImTf$_2$N-PAA) was easily peeled off from the glass plate. Note that the surfaces facing aqueous ammonia and the glass plate are denoted as TOP and BOTTOM surfaces, respectively (Figure S2, supporting information). During the soaking step, water and ammonia molecules diffused into the film from the top surface and triggered the electrostatic complexation between PCMVImTf$_2$N and PAA, a novel self-assembly mechanism we discovered recently for fabricating nanoporous membranes.[13a] As-prepared PCMVImTf$_2$N-PAA membranes feature a combination of porous morphology (Figure S3, supporting information) and a gradient in electrostatic complexation (Figure S4 ~ S5, supporting information) between cationic PCMVImTf$_2$N and the anionic PAA (neutralized by ammonia) from top to bottom.

The PCMVImTf$_2$N-PAA membrane actuator is straight and flat in water (**Figure 1a**, top), and its top surface steadily bends inward (curvature = 0.076 mm$^{-1}$, Figure S6) upon adding acetone up to 0.25 mol% relative to water. By increasing acetone content to 1.5 mol%, the membrane arch bends continuously and ends up with a closed loop (Figure 1a, left column). By decreasing the acetone concentration, the membrane actuator gradually reverts to the original shapes with high accuracy (Figure 1a, right column). Figure 1b quantitatively shows that the curvature of the actuator appears linearly proportional to acetone content, plus being highly reversible. This bending-stretching cycle can be repeated at least 20 times with high accuracy (Figure S7, supporting information). Furthermore the bending kinetics were studied (Figure 1c). Transferring the membrane directly from water into a 1.5 mol% acetone-water mixture, its curvature increases rapidly versus time, then levels off and reaches a plateau after 50 s. Pulling back in water, the recovery of the membrane curvature is slower, which is understandable given the slower rate of acetone releasing from the membrane due to the solvent-polymer attractive interaction. In addition, the temperature influence on the bending actuation was also observed but rather as a secondary effect (Figure S8, supporting information). As both mechanical properties and the ionic bonding in solution are affected by temperature, it remains yet unclear which dominates the temperature effect in the shape deformation of the porous actuators.

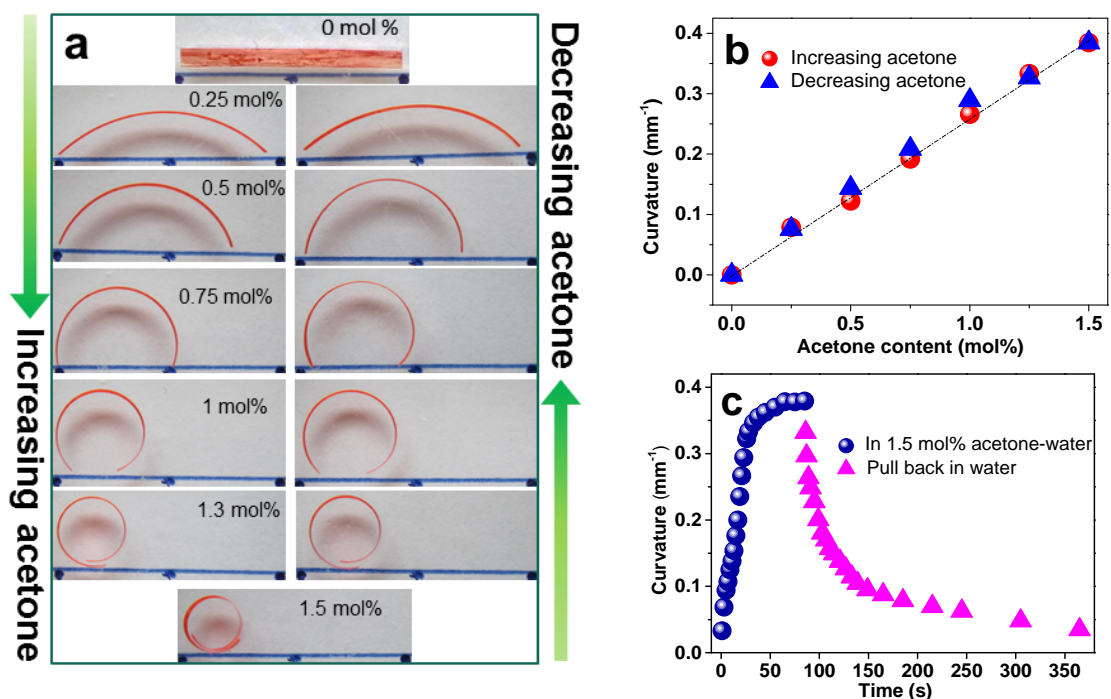

**Figure 1.** (a) shape deformation of a PCMVImTf$_2$N-PAA membrane (1 mm × 20 mm × 30 μm) in response to the molar amount of acetone molecules in a water-acetone mixture at 20 °C (left column: increasing acetone content, right column: decreasing acetone content, the membrane at the top (0 mol%) was a top view and the rest pictures were a side view); note: inserted numbers are the acetone content; the membrane was kept for 5 mins to reach bending equilibrium at each acetone concentrations; the membrane (originally yellowish) was painted in red color for better visibility: the color painting does not affect the actuation; the thick blue bar is a 20 mm scale bar (please see Figure S9, supporting information for the experimental setup), (b) plot of curvature (mm$^{-1}$) of the membrane actuator against acetone content (mol%); (c) plot of curvature (mm$^{-1}$) of the membrane actuator against time when placing it in a 1.5 mol% acetone-water mixture and back in water.

Moreover, we found that the actuator can also respond to other solvents such as tetrahydrofuran, 1,4-dioxane, and ethanol with similarly high sensitivity (Figure S10, supporting information). In the literature, a variety of SSPAs can respond to solvents exchange, but with much lower sensitivity. Here the PCMVImTf$_2$N-PAA actuator is at least one order of magnitude more sensitive than systems recently reported. For example, 50 mol% acetone is required for a phenol-formaldehyde bi-layer film bending to a curvature of 0.35 mm$^{-1}$,[12d] whereas the PCMVImTf$_2$N-PAA actuator bends to a similar curvature (0.38 mm$^{-1}$) requiring only 1.5 mol% acetone. Because it is known that bending actuation is inversely proportional to membrane thickness (Figure S11, supporting information), we also plot "normalized sensitivity" (curvature multiplied by thickness) as well as the apparent sensitivity to compare our membrane's sensitivity with other data reported recently in literature (Table S1, supporting information).

The "stimulus ~ actuator" interaction is commonly recognized as a precondition for SSPAs. In this context a strong "acetone ~ PCMVImTf$_2$N" interaction indeed exists, as hinted by the fact that PCMVImTf$_2$N is soluble in acetone but not in water. Moreover, the membrane possesses a gradient in complexion degree through its cross-section (Figure S4 ~ S5, supporting information), with highest polarity and minimal cross-linking density at the bottom of the membrane. The resultant acetone absorption gradient leads to a swelling gradient across the membrane, decreasing from bottom to top, which in turn drives the bending of the membrane.[13b] The gradient in cross-linking is also likely to produce a gradient in the elastic modulus through the membrane, however this modulus gradient is unlikely to have a strong effect on bending compared to the role of the swelling gradient, as known from the classical analysis of bi-metal thermostats.[17] One potential advantage of a graded membrane as opposed to classical bi-layers is the reduction in interfacial stresses that are produced in a bi-layer. Such stress concentrations are unwanted as they may increase the likelihood of delamination and failure in a bi-layer, but also represent stored elastic energy (coming from the energy of the solvent) that serves no actuation function. By creating appropriate gradients in swellability one may improve efficiency giving actuation at lower acetone concentrations.

In addition, the PCMVImTf$_2$N-PAA actuator membrane is nanoporous (30-100 nm in pore size), in stark contrast to common SSPAs that are dense.[12] The pore channels not only accelerate mass transport of solvents into the membrane, but also weakens the overall bending rigidity, since part of its solid bulk (where the pores stay) is replaced by mobile liquids (Figure 2a). Put another way, pore structures circumvent the need for a high acetone concentration normally required for driving the molecular diffusion and penetration in dense materials, thus leading to a higher sensitivity to solvent concentration.

Reference experiments support our mechanistic views. First, the physical blend of PCMVImTf$_2$N and PAA as control membranes WITHOUT electrostatic complexation and pore structure show negligible response to acetone solvent (Figure S12, supporting information), verifying that the combination of electrostatic complexation and porous architecture are prerequisites for sensitive actuation. Additionally, by modulating the molecular weight of PAA, the membranes were tailored from highly nanoporous to less porous and finally non-porous states (Figure 2b, bottom panel). Consequently, the less porous actuator (Figure 2b, sample on the right) shows much smaller bending in a 1.5 mol% acetone-water mixture than porous films. Given the same chemical nature of the three membranes, unambiguously the pores are playing critical roles improving their sensitivity. However, because the membrane's bending actuation is affected by a multiple of pore structural parameters, future study is needed to engineer these structure features, such as pore size, pore size distribution and pore shape, for task-specific actuation.

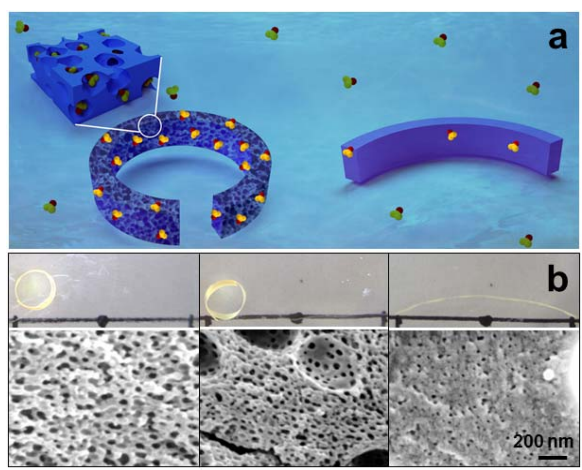

**Figure 2.** (a) A schematic mechanism of the sensitive actuation of a porous membrane actuator (on the left) compared to dense one (right), (b) effect of PAA's molecular weight on the bending actuation of three PCMVImTf$_2$N-PAA membranes (1 mm × 20 mm × 30 μm) placing in 1.5 mol % acetone-water mixture at 20 °C for 10 mins (top panel, NOT painted in red color), and their corresponding porous architectures (bottom panel, SEM pictures of membrane cross-sections). Note: the three membranes were made in the same procedure, except that the molecular weight of PAA from the left to right is 2 000, 5 000, and 100 000 Da, respectively.

The actuator's high sensitivity allows for functionality unattainable with common SSPAs, such as discriminating solvent quality even including isomers. In butanol isomer-water mixtures it shows different bending curvatures (red bars in Figure 3a and inset pictures). To the best of our knowledge this represents the first trial of "reading" solvent isomers, i.e. subtle solvent quality, by SSPAs. We conducted molecular dynamics (MD) simulations to shed more light on the molecular interactions and adsorption processes between the polymer membrane and the butanol isomers. Since the preferential adsorption of solvent to the ionic-liquid-like (IL) groups is responsible for the outstanding membrane properties (sensitivity and selectivity), we simulated the solvation of one PIL-ion pair in the presence of different solvent compositions (see section 3, supporting information). We indeed find an excess adsorption (over water) to the PIL-ion-pair for all three butanol isomers. We have quantified the adsorption with the common adsorption coefficient Γ which has positive values if the butanol is in excess over water; see the blue bars in Figure 3a. Hence, all butanol isomers solvate the PIL-ion-pair better than water and will lead to a larger swelling (and bending) of the membrane due to an enhanced osmotic pressure. Moreover, we see in Figure 3a that the butanol isomers feature an increasing adsorption in the order 1-butanol > 2-butanol > isobutanol. Our mechanistic view is fully supported by the experimental fact that the curvature of the actuator follows exactly the same trend if solvated by these isomers. From the MD simulations we find that on a microscopic level the adsorption trend is related to subtle changes of the detailed interaction of the isomers (with varying hydrophobicity) to the PIL-ion pairs, see Figure 3b for representative simulations snapshots and cartoons depicting the molecular structure. More details to the simulation results can be found in the supporting information (Figure S13 ~ S18).

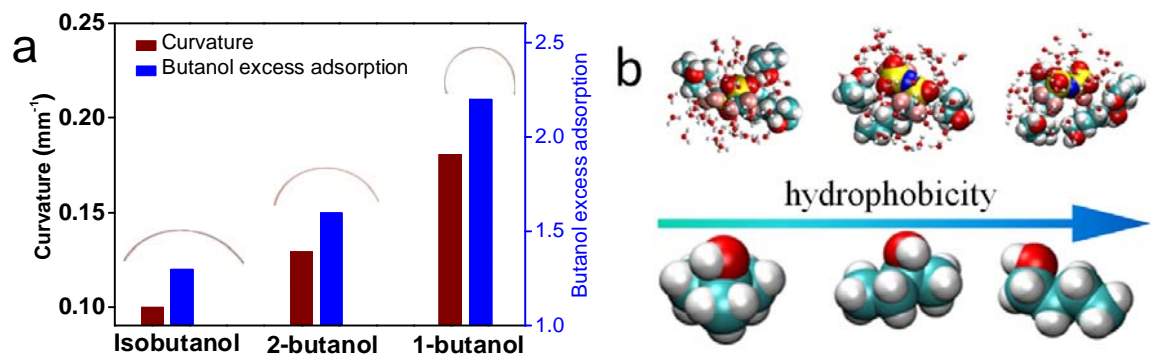

**Figure 3.** (a) Red column: curvature of PCMVImTf$_2$N-PAA actuator membrane in water containing 1.25 mol% butanol isomers (isobutanol, 2-butanol, and 1-butanol) at 20 $^o$C. Blue column: excess absorption of butanol isomers to the PIL-ion pair calculated by all-atom molecular dynamics computer simulations. The inserts are photographs of the bent membrane arch (top view); (b) representative scheme showing the composition of the 1$^{st}$ solvation layer of PIL-ion pair for the three isomers (top panel) and the growing hydrophobicity (and decreasing solubility) of the respective butanol isomers (bottom panel, isobutanol, 2-butanol, and 1-butanol from left to right).

In addition the actuator is combined with other beneficial functionalities such as cooperative actuation, i.e., a group of individual actuators could work cooperatively to accomplish more complicated tasks. This feature is viable even at a relatively small signal input owing to the actuator's high sensitivity. For example, 30 pieces of PCMVImTf$_2$N-PAA membranes were put in a 5 mol% acetone-water mixture, which simultaneously bent and interpenetrated into each other, forming a compact "membrane coil" comprised of entangled and interlocked membrane stripes (Figure 4, Figure S19, supporting information). Put back into water, the "membrane coil" dissolves into the original individual membrane shapes (Figure 4). Here the actuator's high sensitivity is required, otherwise the interpenetration and entangling of different membranes is not effective enough to lock the compact "membrane coil". Given that disentangling this "membrane coil" by hand only ends up with membrane rupture, the cooperative actuation hints to micro-devices capable of multistep manipulation and/or fabrications.

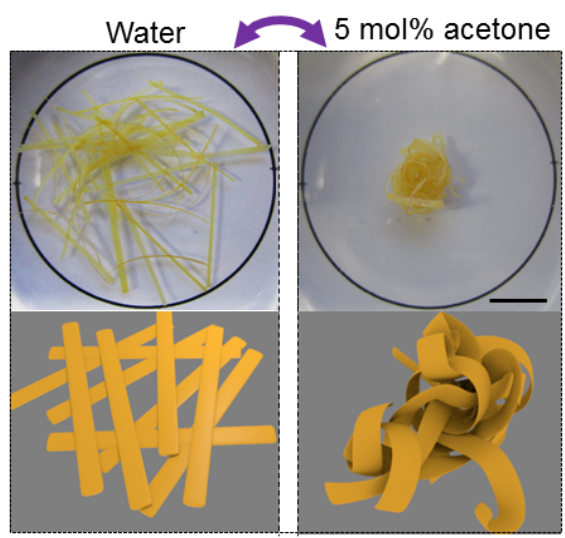

**Figure 4.** Cooperative actuation of 30 PCMVImTf$_2$N-PAA membranes (1 mm × 25 mm × 30 μm) shuttled between water (left) and a 5 mol% acetone-water mixture (right); pictures were taken at a top view; the schemes (bottom panel) illustrate the entangling-dissentangling of membrane stripes. The scale bar is 1 cm.

In summary, we introduced a new concept for fabricating solvent stimulus polymer actuators with unprecedented sensitivity and accuracy. This was accomplished by integrating porous architectures and electrostatic complexation gradients in a poly(ionic liquid) membrane that bears ionic liquid species for solvent sorption. In contact with 1.5 mol% of acetone molecules in water, the actuator membrane (1 mm × 20 mm × 30 μm) bent into a closed loop. While the interaction between solvents and the polymer drives the actuation, the continuous gradient in complexation degree combined with the porous architecture optimizes the actuation, giving it a high sensitivity and even the ability to discriminate butanol solvent isomers. The membrane is also capable of cooperative actuation. The design concept is easy to implement and applicable to other polyelectrolyte systems, which substantially underpins their potentials in smart and sensitive signaling microrobotics/devices.

**Supporting information**

Supporting Information (mateirals characterization, actuation, and molecular dynamic simulations) is available from the Wiley Online Library or from the author.

**Acknowledgements**

QZ, KT and JY would like to thank the Max Planck Society for financial support. JH and JD thank the Alexander-von-Humboldt (AvH) Stiftung, Germany, for financial support. This research is partially supported by the ERC (European Research Council) Starting Grant with project number 639720 – NAPOLI.

DOI: 10.1002/adma.201500533

**The table of contents entry**

**Ultra-sensitive** polymer actuators were prepared which could sense a very small amount (0.25 mol%) of acetone solvent in water, discriminate butanol isomers, and perform cooperative actuation.

**Sensing solvents with ultra-sensitive porous poly(ionic liquid) actuators**

Qiang Zhao,[a] Jan Heyda,[b,e] Joachim Dzubiella,[b,c] Karoline Täuber,[a] John W.C. Dunlop[d] and Jiayin Yuan*[a]

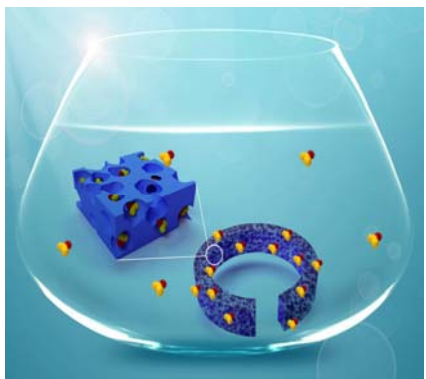
ToC figure

DOI: 10.1002/adma.201500533

# Supporting Information

**Sensing solvents with ultra-sensitive porous poly(ionic liquid) actuators**

Qiang Zhao,[a] Jan Heyda,[b,e] Joachim Dzubiella,[b,c] Karoline Täuber,[a] John W.C. Dunlop[d] and Jiayin Yuan*[a]

## 1. Chemicals, polymers and characterizations

Poly(acrylic acid) (PAA, solid powder, MW=2000 Da), lithium bis(trifluoro methanesulfonyl)imide (LiTf$_2$N, 99.95%), aqueous ammonia (28 w%) were purchased from Sigma-Aldrich and used without further purification. All organic solvents were of analytic grade.

FT-IR spectra were performed on a BioRad 6000 FT-IR spectrometer; samples were measured in solid state using a SingleReflection Diamond ATR. Scanning electron microscopy (SEM) was performed on a GEMINI LEO 1550 microscope at 3 kV. Samples were coated with a thin layer of gold before examination. Element (sulfur) analysis of the membrane cross-section was measured by means of EDX (Oxford instruments) via scanning electron microscopy (DSM 940A, Carl Zeiss AG). Poly(3-cyanomethyl-1-vinylimidazolium bis(trifluoromethanesulfonyl)imide), PCMVImTf$_2$N, was synthesized via the method in our previous study (*Chem. Mater.* 2010, 22, 5003–5012), and characterized by the proton nuclear magnetic resonance ($^1$H-NMR, **Figure S1**). Its apparent molecular weight and PDI were 1.15 × 10$^5$ g/mol and 2.95, respectively.

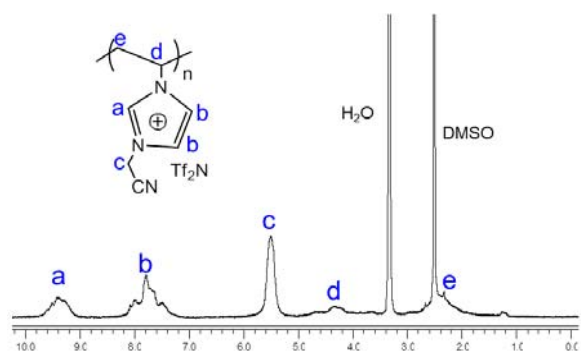

**Figure S1.** Chemical structures and $^1$H-NMR spectra of poly(3-cyanomethyl-1-vinylimidazolium bis(trifluoromethanesulfonyl)imide), PCMVImTf$_2$N.

## 2. Preparation and characterization of PCMVImTf$_2$N-PAA membranes

The actuator membranes were prepared in a method reported previously (**Figure S2**). Typically, PCMVImTf$_2$N (1.0 g) and PAA (0.18 g) were dissolved in 10 mL of dimethylformamide (DMF) solvent, forming a homogeneous solution. Then the solution was cast onto a clean glass plate, dried at 80 $^o$C for 1h, and soaked in aqueous ammonia (0.2 wt%, 20 $^o$C, 2 h). After the soaking step, a free-standing membrane was easily peeled off from the glass substrate (denoted as PCMVImTf$_2$N-PAA). The as-prepared membrane features a nanoporous structure (**Figure S3**). As a result of the pore formation, not surprisingly there is an increase (35 ± 5 %) in membrane thickness after it was treated by aqueous ammonia.

Moreover, when soaking in aqueous ammonia, the ammonia molecules will diffuse into the membrane, and deprotonate the COOH groups on PAA into COO$^-$NH$_4^+$, thus triggering its electrostatic complexation with cationic PCMVImTf$_2$N polymers. As such, the membrane was found to be stable (not to be dissolved) in almost all common organic solvents, indicative of the effective cross-linking via electrostatic interaction.

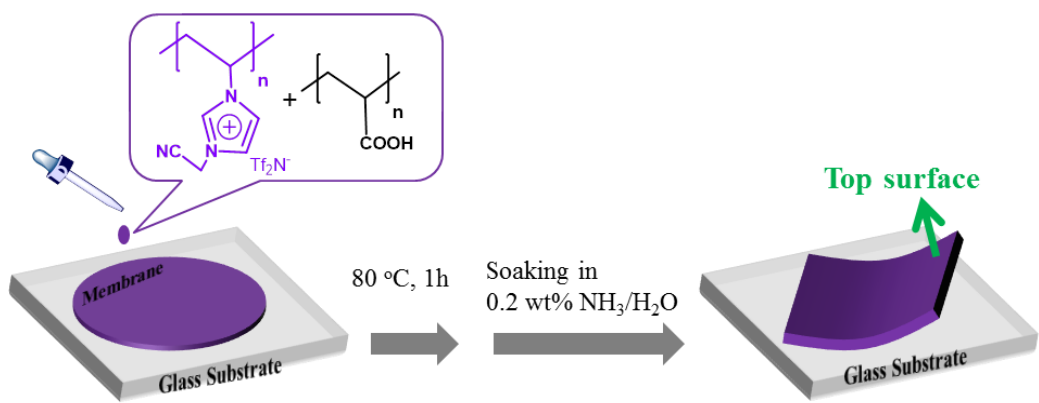

**Figure S2.** Schematic preparation of a PCMVImTf$_2$N-PAA membrane actuator. Note: the membrane surfaces facing the aqueous ammonia and glass plate during the soaking step are denoted as TOP and BOTTOM surfaces, respectively.

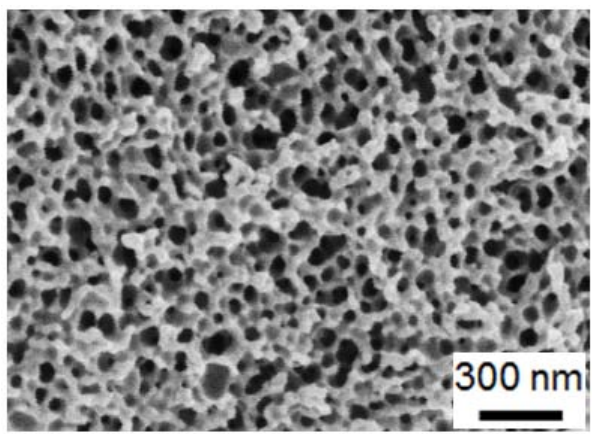

**Figure S3.** Cross-sectional SEM morphology of a PCMVImTf$_2$N-PAA membrane.

The degree of electrostatic complexion (DEC) of the membrane is defined as the ratio of the imidazolium units that ***undergo electrostatic complexation with COO$^-$NH$_4^+$ groups*** to the overall number of imidazolium units (**Figure S4**).

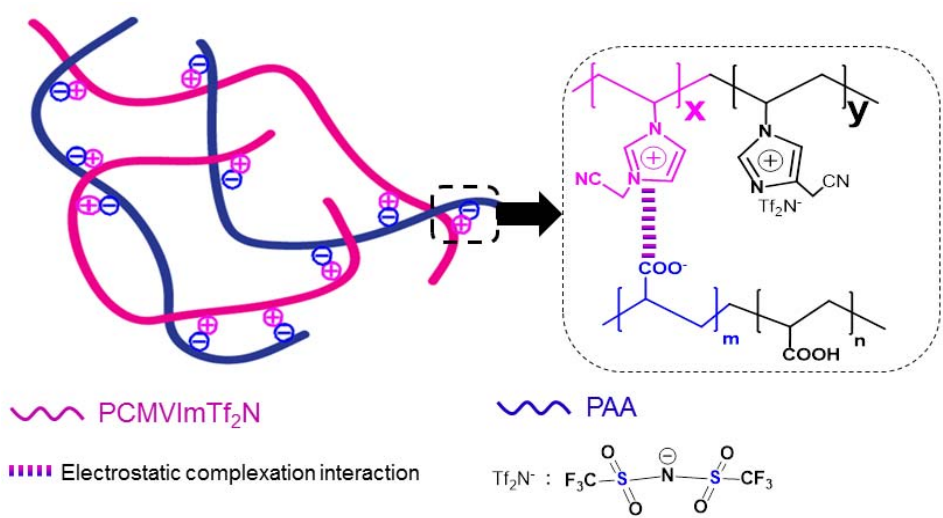

$$\text{DEC}=X/(X+Y) \qquad (1)$$

$$\text{DEC}=(280-486S)/(280-280S) \quad (S: \text{sulfur weight content}) \qquad (2)$$

**Figure S4.** (top) A scheme for defining the degree of electrostatic complexation (DEC) of the membrane; (bottom) equations for defining (eq. 1) and calculating (eq. 2) DEC.

Thus, DEC is expressed in equation (1); whereas, *X* denotes the imidazolium units that undergo electrostatic complexation with COO$^-$NH$_4^+$ groups on PAA; *Y* denotes the

DOI: 10.1002/adma.201500533

imidazolium units that are **NOT** involved in the electrostatic complexation. Note that the electrostatic complexation is accompanied by the release of [$Tf_2N^-$] anions. Thus the value of DEC can be calculated from the content of sulfur element because sulfur exists only in [$Tf_2N^-$] counter anion. As such DEC values at different locations of the membrane are experimentally determined by equation (2), in which *S* is the sulfur content at different locations of the membrane cross-section measured by EDX. **Figure S5** shows the sulfur content along the membrane cross-section; indicating that the DEC decreases with the top-down depth along the cross-section. This DEC gradient is consistent with the membrane formation mechanism. Ammonia diffuses into the membrane from the top surface (membrane-liquid interface) and deprotonates the COOH groups on PAA into carboxylate groups ($COO^-NH_4^+$), thus the DEC is higher at the places closer to the top surfaces.

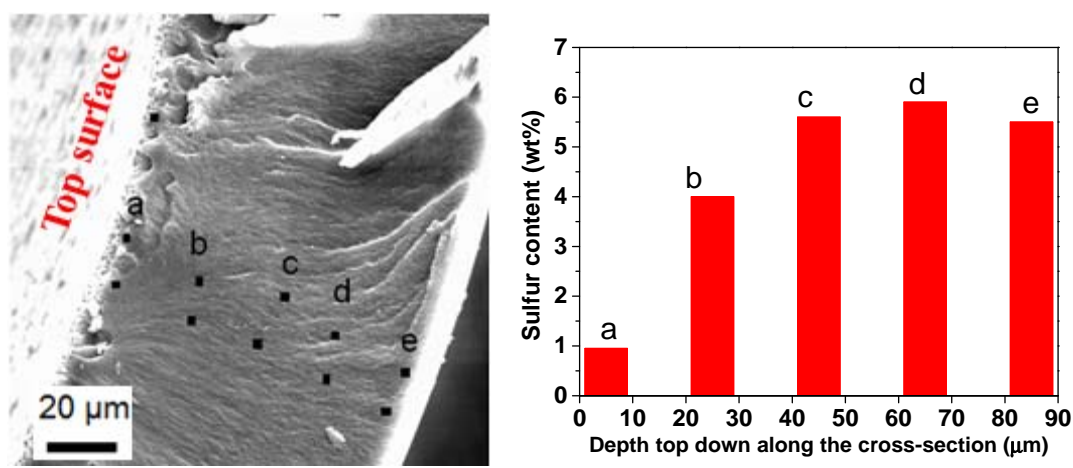

**Figure S5.** Sulfur content at different locations along the cross-section of the membrane actuator from the top to the bottom. On the left is the SEM picture of the membrane cross-section and the positions (a - e) taken for element analysis. Please note: here the membrane sample for measuring the cross-sectional compositions is ca. 90 μm thick. This membrane is thicker than membranes for actuation experiments (ca. 30μm), so that more positions along the cross-section can be tested.

3. **Actuation of PCMVImTf2N-PAA membrane actuators.**

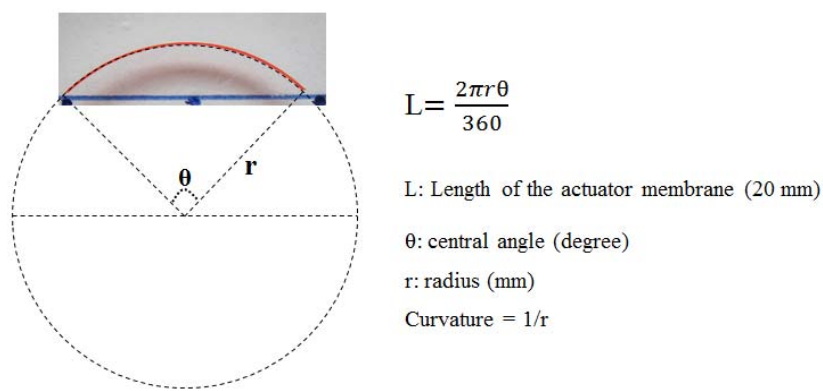

**Figure S6.** Measuring the central angle (θ, degree) of PCMVImTf$_2$N-PAA membrane (1 mm × 20 mm × 30 μm) placed in 0.25 mol% acetone-water mixture at 20 °C. Note: curvature was calculated through equations on the right side. Curvatures of all membrane arches in Figure 1 were obtained in the same way.

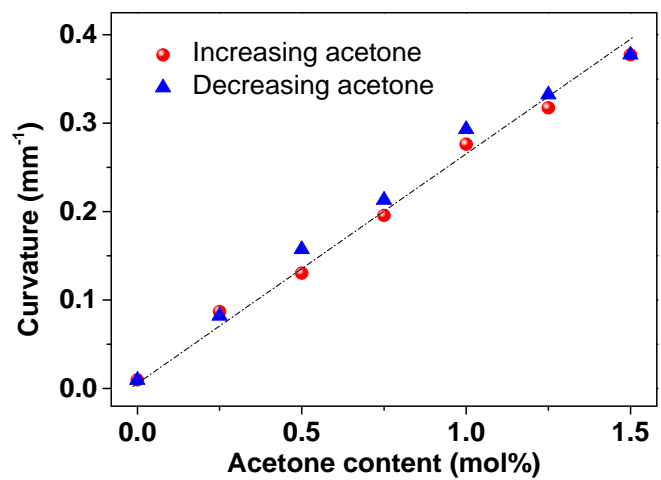

**Figure S7.** Curvature of a PCMVImTf$_2$N-PAA membrane actuator versus acetone content after 20 times cycling, for increasing (red) and decreasing (blue) acetone content from 0 to 1.5 mol%.

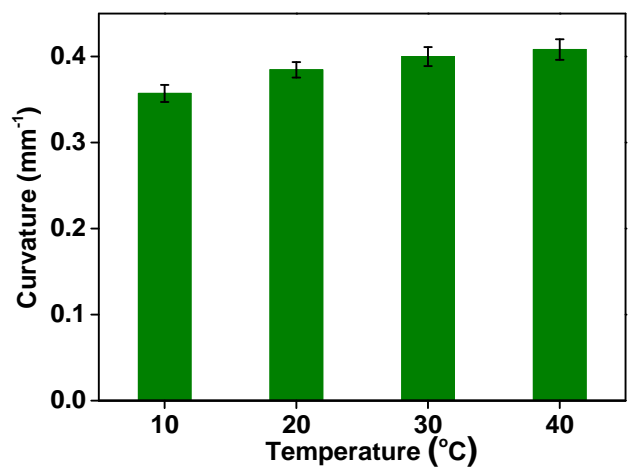

**Figure S8.** Variation of curvature of a PCMVImTf$_2$N-PAA membrane (1 mm × 20 mm × 30 µm) in 1.5 mol% acetone-water mixture at different temperatures.

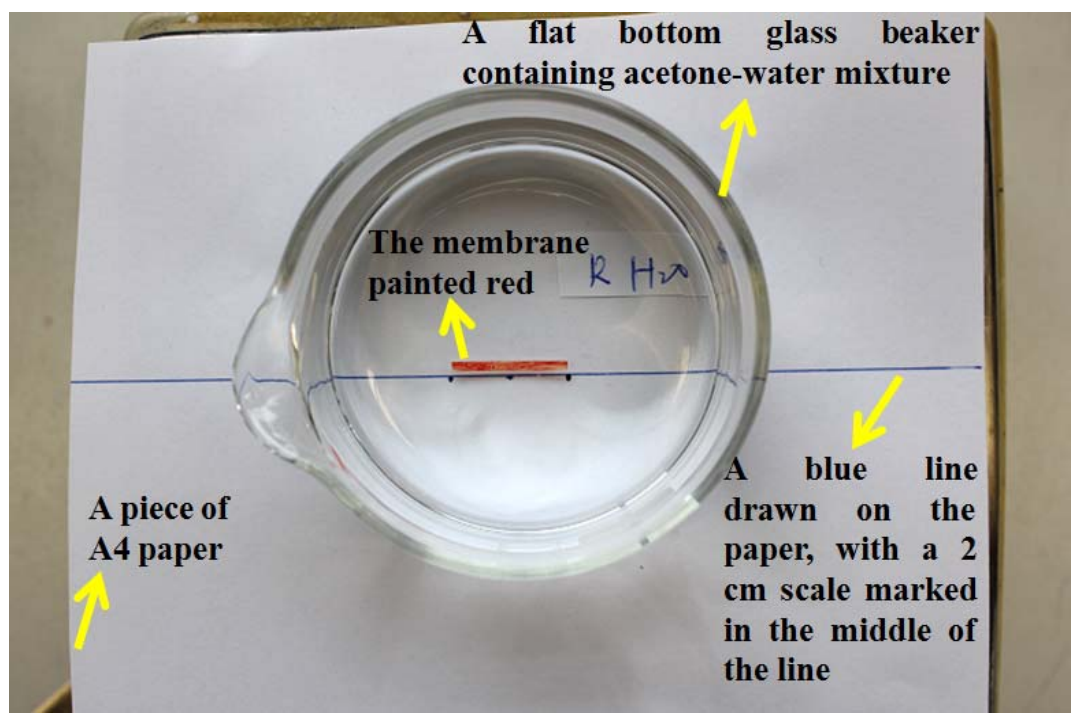

**Figure S9.** A photograph (top view) of the experimental setup for Figure 1a. Please note: the glass beaker was sealed during the experiment.

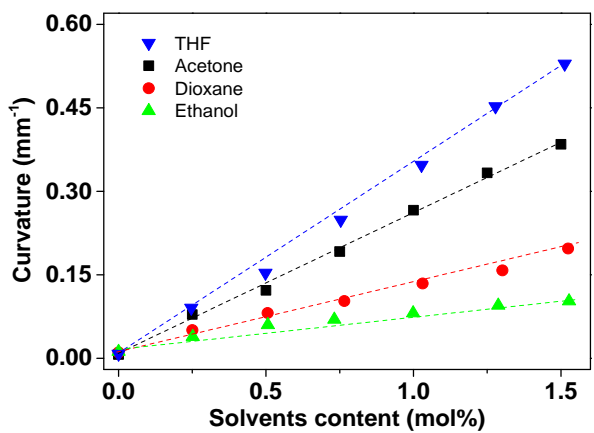

**Figure S10.** Effect of solvents (THF, acetone, dioxane, ethanol) content on curvature of PCMVImTf2N-PAA membrane actuator (1 mm × 20 mm × 30 μm), respectively.

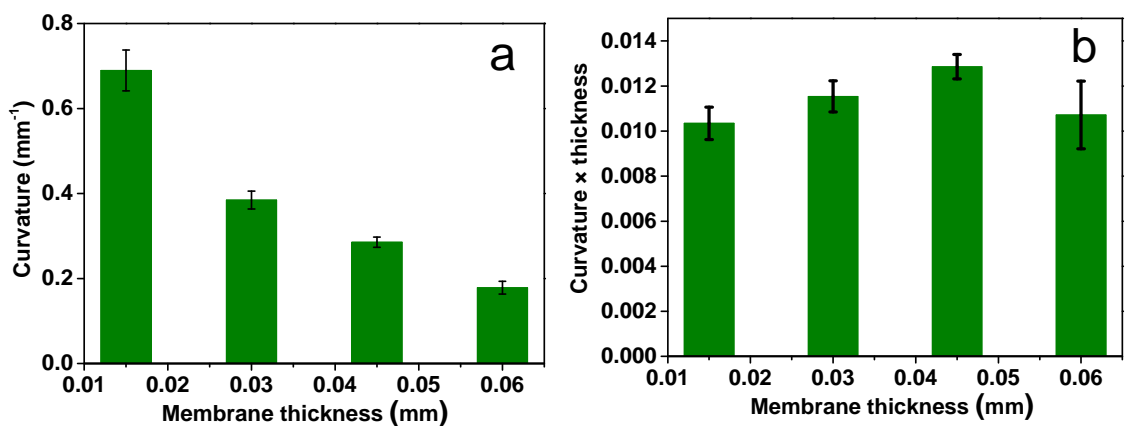

**Figure S11.** Effect of thickness of the PCMVImTf2N-PAA membrane actuator (1 mm × 20 mm) on its bending curvature (a) and normalized curvature (b) in 1.5 mol% acetone-water mixture at 20 °C.

**Table S1.** Comparison of the sensitivity of PCMVImTf$_2$N-PAA actuator with relevant literature data published recently.

| Materials | Organic solvents | Membrane thickness (mm) | Apparent Sensitivity [a] | Normalized Sensitivity [b] | Ref |
|---|---|---|---|---|---|
| Porous PCMVImTf$_2$N-PAA films | Acetone | 0.03 | 0.256 | 7.68 × 10$^{-3}$ | This work |
| Bilayer phenol-formaldehyde | Acetone | 0.045 | 0.0071 | 3.19 × 10$^{-4}$ | [S1] |
| films. | | | | | |
| PTMSDPA/BOPP bilayer films [c] | Acetone | 0.0245 | -0.017 | $-4.17 \times 10^{-4}$ | [S2] |
| Poly(nOBA/C6M)$^-$K$^+$ films | Acetone | 0.018 | 0.0042 | $7.56 \times 10^{-4}$ | [S3] |

a: Apparent sensitivity is defined as the change of curvature achieved by adding 1 mol% organic solvents, i.e., the amount of curvature change ($\triangle c$) divided by the amount of added solvent (S, mol%): $\triangle c/S$. Please note: positive value of sensitivity means the membrane is straight in water and bends upon adding organic solvents; negative value of sensitivity means the membrane is curved in water and unbends upon adding organic solvents.

b: Normalized sensitivity equals apparent sensitivity multiplied by thickness.

c: PTMSDPA: poly[1-phenyl-2-( p -trimethylsilyl)phenylacetylene]; BOPP: biaxially oriented polypropylene

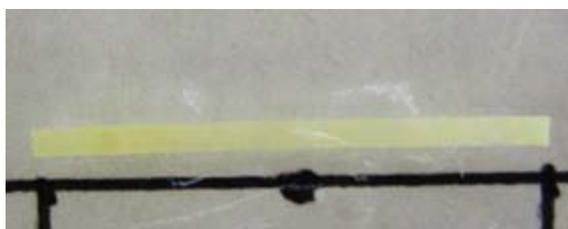

**Figure S12.** An optical photograph of a piece of PCMVImTf$_2$N-PAA membrane (dense structure, without electrostatic complexation) placed in 1.5 mol% acetone-water mixture at 20 $^o$C. Note: the membrane in Figure S12 was not painted with red color and shows its original yellowish color.

4. **Molecular dynamics (MD) computer simulations of PIL-ion pair in water-butanol mixtures.**

Consistently with our recent publication,[S4] a single pair of the ionic liquid (IL) cation [3-cyanomethyl-1-vinylimidazolium] (CMVIm)] and anion bis(trifluoromethanesulfonyl)imide (bistriflate anion, Tf$_2$N) in water mixed with butanol was investigated in terms of all-atom MD simulation. The preferential interaction between the PIL-ion pair and three butanol isomers were individually calculated and compared, namely: 1-butanol (IUPAC: but-1-ol), 2-butanol (but-2-ol), terc-butanol (2-methyl-prop-2-ol). The simulations were performed with the Gromacs 4.5.3 simulation package[S5] using the all-atom OPLS (nonpolarizable) force field.

The force field of the Tf$_2$N anion and imidazolium-based IL was described as put forward recently.[S6] The *1*- and *terc*-butanol force fields were taken from a recently developed solvent database,[S7] which uses the OPLSAA atom types. The force field for 2-butanol was not available, therefore it was built from force-field of 1-butanol, since both alcohols possesses the same atom types. The partial charges were distributed on all butanol isomers in the same way: H(C)=0.06e, C(H)=-n*0.06e (n is the number of covalently bound hydrogens to carbon atom), O(H)=-0.683e, H(O)=0.418e, and partial charge on C(OH) was determined so that the whole alcohol molecule is electroneutral.

The simulations were performed in the constant temperature and constant pressure Gibbs ensemble. Temperature and pressure at ambient conditions (300 K and 101 kPa) were controlled by a weak velocity rescale coupling scheme,[S8] and the Parinello-Rahmann barostat,[S9] respectively. Electrostatic interactions were properly calculated by the particle mesh Ewald summation with standard cut-off and grid parameters.[S6] The integration time step in the simulations was 2 fs and we gathered statistics every 1 ps. We first performed a 20 ns equilibration phase of all systems, which was followed by an at least 80 ns of production phase, in order to obtain properly converged radial distribution functions for ions and solute. To avoid finite size effects in solution structure, we studied relatively large systems with an equilibrium length of approx. 4.5 nm of the cubic and periodically repeated box. Here, a single cation-anion pair was immersed in 2760 SPC/E water molecules,[S10] and 30 or 60 butanol molecules, giving rise to a butanol molar density of 0.6M, and 1.2M respectively.

From the solution structure, the solvation properties, in particular the excess adsorption of butanol over water (i.e. replacement of solvent by cosolvent), can be obtained as detailed in the following.

### 3.1. Analysis of simulation data

From the simulations we calculated the average structure of water (W) and butanol (S) around the PIL-cation/anion (+ and -) in terms of the radial pair distribution functions between the ions (+,-) and the solvents (W,S) $g_{-W}(r)$, $g_{+W}(r)$, $g_{-S}(r)$, $g_{+S}(r)$. The results gathered for two concentrations were compared and evaluated.

The water and cosolvent adsorption (with respect to an ideal gas), also known as excess coordination number, $N_{ij}^{ex}$,[S11] is then defined by volume integration over the PIL-cation/anion-water ($g_{iW}(r)-1$) and PIL-cation/anion-cosolvent ($g_{iS}(r)-1$) structure, multiplied by the density $\rho_W$ and $\rho_S$, respectively.

$$N_{iS}^{ex} = 4\pi\rho_S \int_0^\infty (g_{iS}(r) - 1)r^2 dr \qquad (S-MD1)$$

$$N_{iW}^{ex} = 4\pi\rho_W \int_0^\infty (g_{iW}(r) - 1)r^2 dr$$

The cosolvent adsorption for the PILion-pair, $N_{iS}^{ex}$, is then obtained as the sum of contribution from cation and anion respectively:

$$N_S^{ex} = N_{+S}^{ex} + N_{-S}^{ex} \qquad (S-MD2),$$

and the water adsorption is obtained analogously.

Another important thermodynamic quantity in water:cosolvent mixtures is the excess adsorption of cosolvent with respect to the water, also known as preferential binding

DOI: 10.1002/adma.201500533

coefficient $\Gamma_{ij}$. This is defined by the volume integration of the solute-solvent and solute-cosolvent structure difference ($g_{iS}(r)$- $g_{iW}(r)$) and multiplied by the cosolvent density $\rho_S$

$$\Gamma_{iS} = 4\pi\rho_S \int_0^\infty (g_{iS}(r) - g_{iW}(r))r^2 dr \qquad (S-MD3)$$

The solvent excess adsorption for the PILion-pair, $\Gamma_S$, is again obtained as the sum

$$\Gamma_S = \Gamma_{+S} + \Gamma_{-S} \qquad (S-MD4).$$

All these thermodynamic parameters are summarized in the **Table S1** and plotted in **Figure S16** with respect to butanol concentration.

### 3.2. Results

Applying MD simulations, we have quantified the structure around the PIL cation and anion in terms of radial distribution functions at two butanol concentrations (0.6 M and 1.2 M). Results of the structure are presented in **Figure S13** and **Figure S14**.

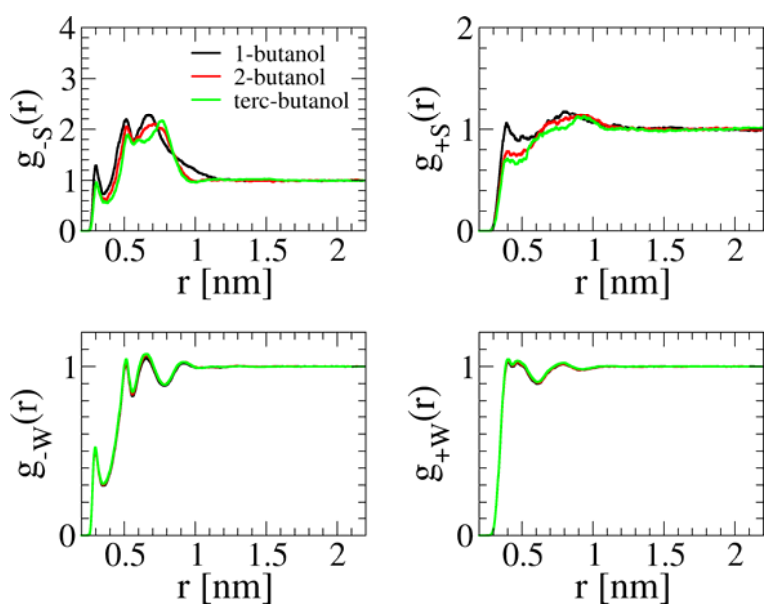

**Figure S13.** Radial distribution function of 0.6 M butanol (S) and water (W) around PIL-anion (-) and PIL-cation (+). It is evident that the butanol is more attracted to the PIL-anion

than to cation, and for both PIL-ions, the **order is 1-butanol > 2-butanol > terc-butanol**. In contrast, the water solution structure around PIL-ions is almost indistinguishable in all three cases.

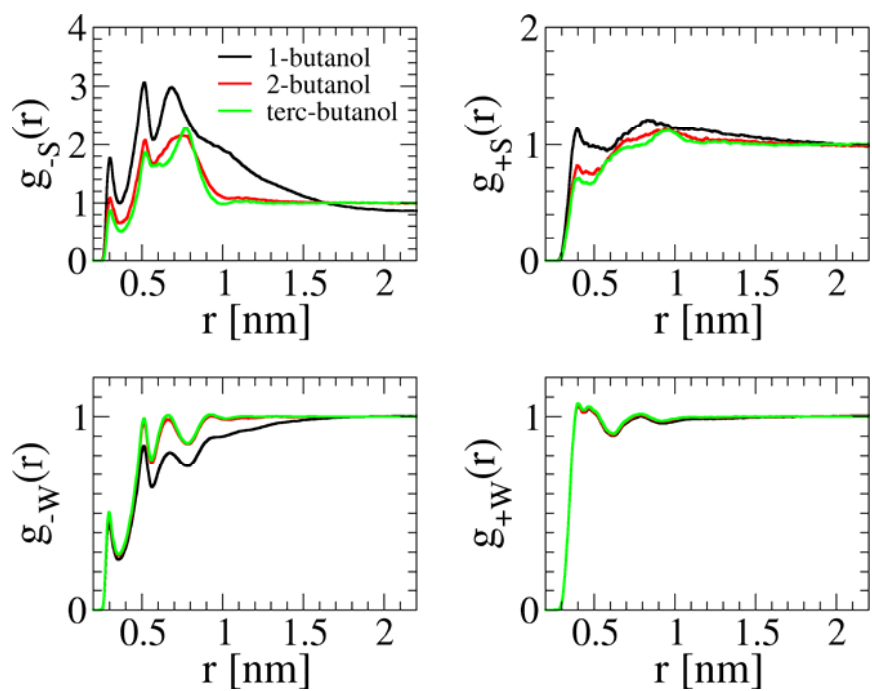

**Figure S14.** Radial distribution function of 1.2 M butanol (S) and water (W) around PIL-anion (-) and PIL-cation (+). It is evident that the butanol is more attracted to the PIL-anion than to cation, and for both PIL-ions, the order is **1-butanol > 2-butanol > terc-butanol**. In contrast, the water solution structure around the PIL-ions is almost indistinguishable in three cases. *Please note that the 1-butanol starts to aggregate at this concentration (it is around its experimental water solubility limit). This is also evident from water-butanol solution structure (data not shown).*

Applying eq. (S-MD1), the water and cosolvent excess adsorption numbers were obtained (see **Figure S15** and the summary in **Table S2**). Positive values can be interpreted as adsorption, negative values as exclusion. We found specific adsorption of the cosolvent (at both concentrations) to the cation-anion pair in the order:

**1-butanol >> 2-butanol > terc-butanol >> water**

This order is in accord with the experimental PIL solubility and curvature data. We note that the butanol adsorption was found always to be higher to the PIL-anion as compared to the PIL-cation.

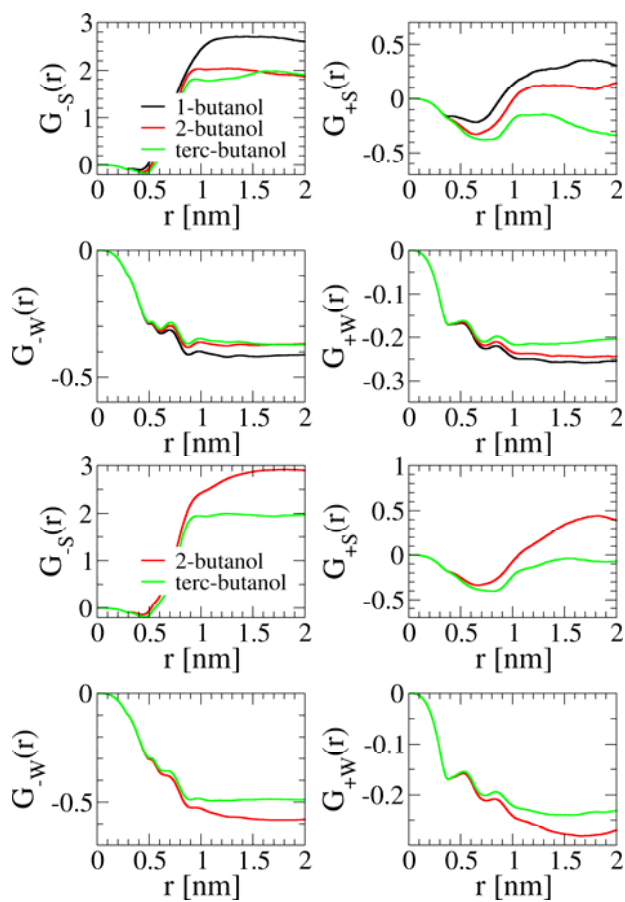

**Figure S15.** Running Kirkwood-Buff integrals ($\rho_j G_{ij} = N_{ij}^{ex}$, i.e. concentration normalized adsorption number) for butanol (S) and water (W) around the PIL-anion (-) and the PIL-cation (+), evaluated by eq. (S-MD1) for the lower butanol concentration (0.6 M) on the left and at the higher butanol concentration (1.2 M) on the right. The thermodynamically relevant value is read from the plateau (i.e. in the region where g(r) reaches unity); here we have chosen the value at r = 1.5 nm. The observed order is 1-butanol >> 2-butanol > terc-butanol for both PIL-

ions. In contrast, the water solution adsorption around PIL-ions is similar and negative in all butanol solutions. Please note that the 1-butanol starts to aggregate at 1.2M concentration, for clarity this data are not presented.

**Table S2.** Summary of the thermodynamic parameters – Kirkwood-Buff integrals $G_{ij}$, Adsorption numbers $N_i^{ex}$, and preferential binding coefficient $\Gamma$, for one PIL-ion-pair in water-butanol solutions. Two butanol concentrations are compared (0.6 M and 1.2 M), and the neat water solution serves as a reference for water adsorption ($N_W^{ex}$). See also **Figure S16**, for a presentation of the concentration dependence of these parameters.

| System | $\rho_{butanol}$ | $G_{PIL-W}$ | $G_{PIL-S}$ | $N_W^{ex}$ | $N_S^{ex}$ | $\Gamma$ |
|---|---|---|---|---|---|---|
| neat water | 0 | -0.49 | ND | -26.7 | ND | ND |
| 1-butanol | 0.6 | -0.66 | 3.15 | -34.7 | 1.8 | 2.2 |
| 1-butanol | 1.2 | ND | ND | ND | ND | ND |
| 2-butanol | 0.6 | -0.62 | 2.15 | -32.6 | 1.2 | 1.6 |
| 2-butanol | 1.2 | -0.88 | 3.1 | -46.2 | 3.5 | 4.5 |
| terc-butanol | 0.6 | -0.59 | 1.75 | -31.0 | 1.0 | 1.3 |
| terc-butanol | 1.2 | -0.74 | 1.9 | -38.9 | 2.2 | 3.0 |

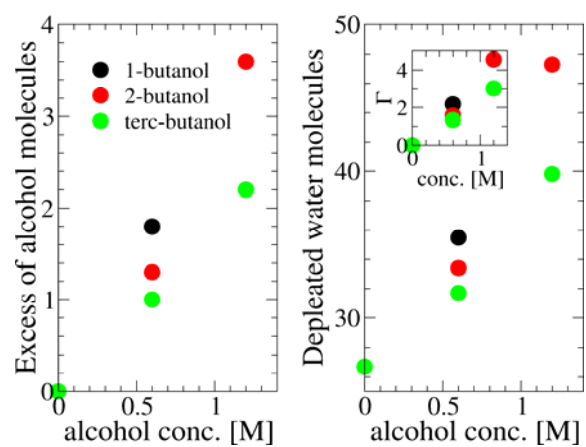

**Figure S16.** Excess of butanol molecules (left) and the exclusion of water molecules (right) from the PIL-ion-pair presented as a function of butanol concentration (see Table S2). The preferential binding coefficient $\Gamma$ (excess of butanol over water), is provided as an inset of the right figure, and the positive value documents the enrichment of PIL-ion-pair vicinity in butanol. The observed order of the butanol excess is **1-butanol >> 2-butanol > terc-butanol**. The same order holds for the water exclusion.

*3.3. Mechanistic interpretation at microscopic level:*

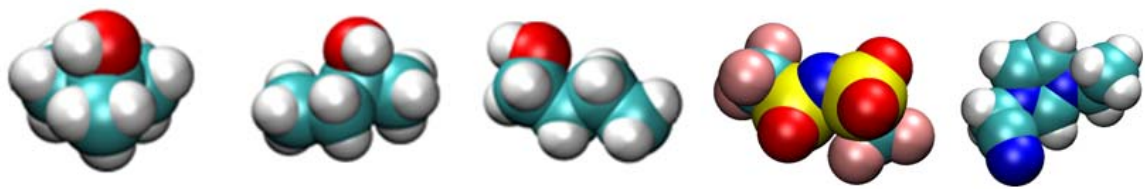

**Figure S17.** Space filling representation of isobutanol, 2-butanol, 1-butanol Tf$_2$N anion, and imidazolium based IL cation (left to right). Colors: H-white, C-cyan, O-red, S-yellow, N-blue, F-pink. All molecules are modelled as flexible in the MD simulation.

**Already from the structure and hydration of alcohol and PIL molecules (see Figure S17), we can raise three points which are likely responsible for the butanol preferential adsorption to PIL ion-pair in aqueous solutions.**

(1) Water hydrates only weakly the hydrophobic regions (fluoro- and hydrocarbon groups) of PIL ions.

(2) Butanol isomers are increasingly hydrophobic 1-butanol > 2-butanol > terc-butanol, i.e., the large hydrophobic part is weakly hydrated, since the effect of the hydroxyl group induces only local hydrophilicity.

(3) Butanol isomers have very limited H-bonding propensities (only a single OH group), and the PIL ions can serve only as H-bonding acceptors (via S=O, and C≡N group).

To sum up, water, as a highly polar medium with dynamic H-bonding networks, is preferably depleted from the PIL vicinity (first 1-2 solvation layers) and replaced by less polar butanol molecules. The effect is more pronounced for the PIL anion, as documented in **Figure S18** on the examples of the solvent/cosolvent distribution for three investigated butanol isomers. The more polar regions of IL, such as -O$_2$S-N-SO$_2$-, stay hydrated.

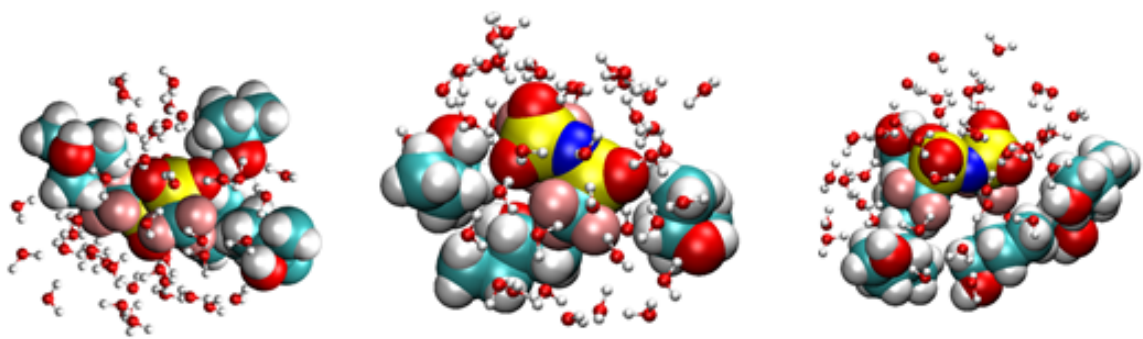

**Figure S18.** Illustrative distribution of butanol molecules (in space filling representation) and of water molecules (spheres-and-sticks) around the bis-triflate anion. From left to right: isobutanol, 2-butanol, and 1-butanol. Note the hydrophobic regions of bis-triflate are solvated by butanol, while polar regions bis-triflate by water.

As clearly visible from **Figure S18**, there is only minor difference between the structural arrangements of the butanol isomers in PIL vicinity. However, based on their different hydrophobicity, ordered from most hydrophobic to least hydrophobic, **1-butanol > 2-butanol > *terc*-butanol**, and the different propensity of butanol isomers to create an H-bond, **1-butanol > 2-butanol > *terc*-butanol** (best H-bond to worst), we suppose that the 1-butanol physico-chemical properties compete best with water for interactions with our PIL-ion-pair. We note that the PIL-cation-anion pair formation is observed to be of transient nature and is quantitatively the same in all three water-butanol solutions (data not shown).

### 3.4. Summary – action of butanol isomers on the actuator

Our MD simulations and subsequent analysis provide clear evidence that 1-butanol better solvates the PIL ion-pair compared to 2-butanol and terc-butanol by preferentially adsorbing mainly to the bis-triflate anion and replacing a large number of less preferably located water molecules. This butanol excess leads to a decrease in interfacial tension and the systems

'opens up' to more favorable butanol adsorption. From a mechanical point the osmotic pressure of the solvent increases and the polymer swells.

## 5. Cooperative actuation of PCMVImTf$_2$N-PAA membrane actuators.

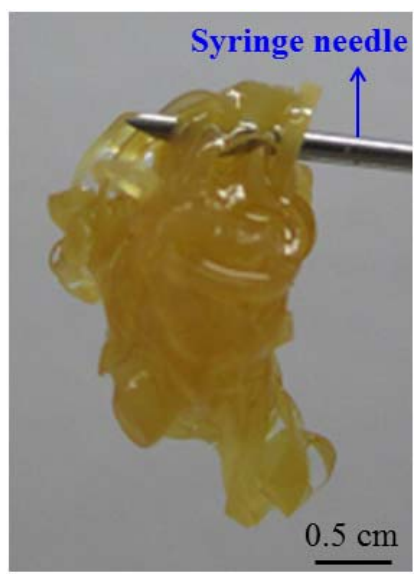

**Figure S19.** A photograph showing that the "membrane coil" in Figure 4 could be suspended in air by poking a syringe needle through it, indicating that the coil is comprised of entangled and interlocked membrane stripes. Note: the membranes in Figure S16 were not painted with red color and show their original yellowish color.

**References to SI**